\documentclass{llncs}

\usepackage[utf8]{inputenc}
\usepackage{times}
\usepackage{amsmath}
\usepackage{amssymb}
\usepackage[all]{xy}
\usepackage[english]{babel}
\usepackage{graphicx}
\usepackage{color}
\usepackage{subfigure}
\usepackage{here}
\usepackage{makeidx}

\newcommand {\dom}{\mathsf{dom}}

\newcommand{\ignore}[1]{}

\newcommand {\A}{{\mathcal A}}
\newcommand {\limit}{\mathsf{limit}}

\begin{document}

\frontmatter          

\pagestyle{headings}  

\addtocmark{} 

\mainmatter            

\title{Deciding the 
Borel complexity of regular tree languages}

\titlerunning{Deciding the Borel complexity  of regular tree languages}  

\author{Alessandro Facchini\thanks{The author is supported by the \emph{Expressiveness of Modal Fixpoint Logics} project realized within the 5/2012 Homing Plus programme of the Foundation for Polish Science, co-financed by the European Union from the Regional Development Fund within the Operational Programme Innovative Economy (``Grants for Innovation'')} \and Henryk Michalewski}
\institute{University of Warsaw \\ 
\email{\{A.Facchini,H.Michalewski\}@mimuw.edu.pl}}

\authorrunning{A. Facchini \and H. Michalewski}   

\maketitle              

\begin{abstract}
\noindent We show that it is decidable whether a given a regular tree language belongs to the class ${\bf \Delta^0_2}$ of the Borel hierarchy, or equivalently whether the Wadge degree of a regular tree language is countable.
\end{abstract}

\section{Introduction}
In \cite{niwalu} there was given an algorithm which for a \emph{deterministic} parity tree automaton $\A$ decides whether the language $L(\A)$ is Borel. This was further extended to a finer classification in \cite{murlak2} and finally to a full Wadge classification in \cite{murlak}. The algorithms look for a pattern in the graph of the automaton and decide the Borel and Wadge classes upon finding of these special patterns. 

Similar problems for \emph{non-deterministic} parity tree automata seem to be much harder. 
Recently in \cite{bp} was provided an algorithm which decides for a given non-deterministic parity tree automaton $\A$, 
whether $L(\A)$ is a Boolean combination of open sets. For other Borel classes there was no known algorithm. This paper provides a relatively simple extension 
of the result in \cite{bp} to the class of ${\bf \Delta^0_2} = {\bf\Sigma^0_2}\cap{\bf\Pi^0_2}$ sets, that is the sets which are simultaneously presentable as countable unions of closed sets and countable intersections of open sets. This result is presented in Section \ref{sec:the main result} in Theorem \ref{theorem:main}. The proofs in \cite{bp} are based on an analysis of an algebraic structure computable from $\A$ and the main result states that the language $L(\A)$ is a Boolean combination of open sets if and only if a certain finite number of algebraic requirements hold. Since the class ${\bf \Delta^0_2}$ is bigger, in order to characterize this class, the set of algebraic requirements must be relaxed. In this paper we show that indeed this is the case. Our proofs closely follow the proofs from \cite{bp} with some necessary adjustments. In particular the crucial concept of the topological cutting game introduced in \cite{bp} is considered in this paper not only in the finite, but also in the infinite case. 

The approach presented in \cite{bp} and in the present paper in a certain sense is a reminiscent of the approach applied to deterministic automata in \cite{murlak2,murlak,niwalu}. Namely, the algebraic structure computed from a given automaton $\A$ induces a graph with edges reflecting the algebraic properties. In the deterministic case it is possible to decide Borel and Wadge classes analyzing patterns in the graph of the automaton, in the present paper we are looking for patterns in the algebraic graph. 

Finally let us mention results which provide information about the set-theoretical complexity of a language accepted by a non-deterministic automaton $\A$ assuming some additional properties of $\A$:
\begin{itemize}
\item Rabin in \cite{rabin} proved, that if $L$ and its complement are accepted by a non-deterministic Büchi tree automata, then $L$ is weakly definable, in particular it is Borel.
\item Recently in \cite{cklvb} it was shown using decidability results about the cost functions, that for a given non-deterministic Büchi tree automaton it is decidable whether the language is weakly definable. 
\item In \cite{fms} the decidability results regarding deterministic automata were lifted to 
a more general context of game automata.
\end{itemize}
This paper consists of four Sections: the Introduction, a preliminary Section 2 introducing automata, set-theoretical and algebraic notations, a Section 3 introducing topological games and linking these games to the Wadge hierarchy and Section 4 containing the main result.


\section{Preliminaries}
\subsubsection*{Trees and contexts.}
Given a finite alphabet $A$, a \emph{tree over $A$} is a partial function $t: \{0,1\}^* \to A$ such that its domain $\dom(t)$ is prefix closed. A node of a tree $t$ is an element $v \in \dom(t)$. A left child of a node $v$ of $t$ is the node $v0$, while its right child is $v1$. A leaf of a tree is a node without children. 
We denote by $T_A$ the family of all  trees over $A$.
A set of trees over $A$ is called a tree language, or simply a language.
A \emph{multi-context} over $A$ is a tree $c$ over $A \cup \{\star\}$, where 
\begin{itemize} 
\item $\star \notin A$, and 
\item $\star$ only labels some leaves of $c$. 
\end{itemize}
A leaf of $c$ labelled by $\star$ is called a \emph{port}. Notice that a multi-context may have infinitely many ports. For a multi-context $c$ and a function $\eta$ mapping each port of $c$ to a tree $t$ over $A$, by $c[\eta]$ we denote the tree given by inserting into every port $x$ a tree $\eta(x)$. When $\eta(x)=t$ for each port $x$, we just write $c[t]$.  We say that a tree $t$ extends a multi-context $c$ if there is a mapping $\eta$ such that $c[\eta]=t$. Given a multi-context $c$ and a language $L$, by $[c]^{-1}L$ we denote the language of trees $t\in L$ extending $c$.
 The class generated by $c$ and all possible mappings $\eta$ is denoted by $c[T_A]$. 
A finite multi-context is called a \emph{prefix}.
A multi-context with only one port is called a \emph{context}. 
\subsubsection*{Topology.}
For a finite alphabet $A$, we equip the class $T_A$ of all trees over $A$ with the prefix topology. That is the basic open sets are sets of the form $p[T_A]$, for a prefix $p$ over $A$, and thus the open sets are of the form $\bigcup_{p \in P}p[T_A]$ for some set $P$ of prefixes. 

The class of Borel tree languages of $T_A$ is the
closure of the class of open sets of $T_A$ with respect to countable unions and
complementations. Given $T_A$, the initial finite levels of the
Borel hierarchy are defined as follows: 
\begin{itemize}
\item ${\bf\Sigma}^0_1(T_A)$ is the class of open subsets of $T_A$, 
\item ${\bf\Pi}^0_n(T_A)$ consists of complements of sets from ${\bf\Sigma}^0_n(T_A)$, 
\item ${\bf\Sigma}^0_{n+1}(T_A)$ consists of countable unions of sets from ${\bf\Pi}^0_n(T_A)$. 
\end{itemize}
A much finer measure of the topological complexity is the \emph{Wadge degree} (see \cite[Chapter 21.E]{kechris}).
If $L \subseteq T_A$ and $M\subseteq T_B$, we say that $L$ is \emph{continuously (or Wadge)
reducible} to $M$, if there exists a continuous function $f: T_A \to T_B$ such that $L=
f^{-1}(M)$. We write $L \leq_W M$ iff $L$ is continuously reducible to $M$.
This pre-ordering is called the {\em Wadge ordering}. If $L \leq_W M$ and $M \leq_W L$, then we write $L
\equiv_W M$. If $L \leq_W M$ but not $M \leq_W L$, then we write $L<_W
M$. The Wadge hierarchy is the partial order induced by $<_W$ on the
equivalence classes given by $\equiv_W$. A language  $L$ is called {\em self dual} if it is equivalent 
to its complement, otherwise it is called {\em non self dual}.

Given a certain family of sets $\mathcal{C}$, we say that $M$ is $\mathcal{C}$-hard if $L
\leq_W M$ for every $L \in \mathcal{C}$. A    $\mathcal{C}$-hard set $L$ is said to be $\mathcal{C}$-complete if 
moreover $L \in \mathcal{C}$.
\subsubsection*{Algebra.}
The Wadge hierarchy of the regular languages of infinite words is well understood thanks to a classification result by K. Wagner (\cite{wagner}). In particular from Wagner's result one can derive an algorithm which decides whether a given regular language of infinite words is a Boolean combination of open sets. Being a Boolean combination of open sets is equivalent to being in ${\bf \Delta^0_2}$ class in the context of regular languages of infinite words. This is not true for the regular languages of infinite trees (see \cite[Section 4.1]{bp} and Proposition \ref{rem:example} in Appendix of the present paper for an analysis of one special case).   A natural algebraic interpretation of Wagner's result can be found in \cite[Theorem V.6.2]{pinperrin}. In the case of languages of infinite trees, the algebraic theory is not yet fully developed. As a general reference may serve papers \cite{blumensath,bp,bojtrees,bojidziaszek}. For details of the approach applied in the present paper refer to \cite[Section 3]{bp}. 

Following the approach presented in \cite{bp}, the family of all trees $T_A$ is divided into finitely many Myhill-Nerode equivalence classes $H_L$. Similarly, there are finitely many equivalence classes $V_L$ of contexts. The same holds for multi-contexts with a fixed number of holes. Starting from an automaton accepting language $L$, one can compute families $H_L$ and $V_L$. The equivalence class of a tree $t$ or a context $v$ is denoted $\alpha_L(t)$, $\alpha_L(v)$, respectively. For a given tree $t$ and contexts $v_1,v_2$ multiplication of contexts and trees $v_1t$, $v_1v_2$ naturally induces multiplication between elements of $H_L$ and $V_L$. Similarly, for a given context $v$ the operation of infinite power $v^\infty = vv\ldots$ induces a mapping from $V_L$ to $H_L$. 

Given a regular language $L$, its \emph{strategy graph} $G_L$ is the pair $(V_L \times H_L, E)$ such that 
$((v,h),(v',h')) \in E$ iff there exists a tree $t$ of type
$h$ such that  $t$ can be decomposed as the concatenation of a context of type $v$ and another tree, and 
each prefix of $t$ can be completed into a context of type $v'$.
We thus say that the strategy graph is \emph{recursive} if there exists a strongly connected component that contains two nodes $(v, h)$ and $(v', h')$ with $h \neq h'$.
For a more formal approach to the strategy graph refer to \cite[Section G]{bp}. We will need the following 
\begin{proposition}[\cite{bp}]
\label{proposition:path to edge}
If there exists a path from $(v,h)$ to $(v',h')$ in $G_L$, then there exists an edge from $(v,h)$ to $(v',h')$.
\end{proposition}


\section{Topological complexity and games}\label{section:games}
\subsubsection*{Topological Games.}
Let $L$ and $M$ be two languages. The {\em Wadge game}
$\mathcal{W}(L, M)$ is an infinite two-player game between Player I and Player II. It is defined as follows. During a play Player I constructs a tree $t$ and Player II a tree $t'$. At the first round Player I plays a root of $t$ and Player II plays a root of $t'$, and at each consecutive round both players add a level
to their corresponding tree (thus either Player adds some child to a leaf or Player signalizes that the node will be also a leaf of the final resulting tree of the play by not adding any children to it).
Player I plays first and Player II is allowed to
skip her turn but not forever.  Player II wins the game iff $t \in L
\Leftrightarrow t' \in M$.  
The game was designed precisely in order to obtain a characterisation of continuous reducibility.
\begin{lemma}[\cite{wadge}]\label{lemma:wadge}
Let $L, M$ be two languages. Then  $L \leq_W M$ iff Player II has a winning strategy in the game $\mathcal{W}(L, M)$.
\end{lemma}
From Borel determinacy (\cite{martin}),
if both $L$ and $M$ are Borel, then $\mathcal{W}(L, M)$ is determined.
The ordering $<_W$ restricted to the Borel sets is
well-founded (see \cite[Theorem 21.15]{kechris}). The \emph{Wadge degree} for sets of finite Borel rank can
be 
defined inductively. First, we remark that since every self dual set $A$ is Wadge equivalent to the disjoint union of a certain non self dual set $B$ and its complement $B^\complement$, it is enough to start associating a Wadge degree only to non self dual sets and say that the Wadge degree of $A$ equals the Wadge degree of $B$. 
For each degree there are exactly three equivalence
classes with the same degree, represented by $L$, $L^\complement$ and
$L^\pm$ --- the disjoint union of $L$ and $L^\complement$. 
Clearly $L, L^\complement <_W L^\pm$ and
$L^\pm$ is self dual.

In \cite{dup1}, J. Duparc showed that for non self dual sets, it is possible to determine its sign, $+$
or $-$, which specifies precisely the $\equiv_W$-class. 
For instance, $\emptyset$ and complete open sets have sign $-$, while the whole space and complete closed  sets have sign $+$.
  All self dual sets by
definition have sign $\pm$. Let $\kappa$ be the length of Wadge hierarchy of Borel sets of finite rank. 
Thus an ordinal $\alpha <\kappa$
determines a $\equiv_W$-class, denoted  $[\alpha]^\epsilon$ for $\epsilon\in \{+,-,\pm\}$. 
In the same paper,  in the context of Wadge degrees, Duparc defined  set-theoretical counterparts of ordinal multiplication by a countable ordinal, and (quasi) exponentiation of base $\omega_1$. 
From now on $[\alpha]^\epsilon$ will also denote the canonical sets of Wadge degree generated with Duparc's operations. We present some details of Duparc's construction in the Appendix. 
\subsubsection*{Cutting games.}
Below we define a family of two-player games of perfect information, called \emph{cutting games}. These games were
introduced in \cite{bp}. For the argument in \cite{bp} the most important was the finite version of the game. In the present paper 
we will consider both infinite and finite versions of the cutting game. 

Let $L_i$ ($i=1,2,\dots$) be languages over the alphabet $A$, and let $p$ be a prefix over the alphabet $A$. 
The \emph{simple cutting game} of length $k$, denoted $\mathcal{H}^p_k(L_1,\ldots,L_k)$ is played by two players, Constrainer and Alternator. 
For each $i \in \{1, \dots, k \}$ the $i$-th round of the game is played as follows:
\begin{itemize}
\item Alternator chooses a tree $t_i \in L_i$ extending the prefix chosen in the previous round by the Constrainer; in the first round of the game Alternator must choose an extension of the given prefix $p$,
\item Constrainer chooses a prefix of the tree $t_i$.
\item If Alternator cannot move, she loses, but if she survives k rounds then she wins.
\end{itemize}
The \emph{infinite cutting game}, denoted by $\mathcal{H}^p_\infty(L_1,\dots)$, is played just like a simple game but without the restriction to a fixed given number of rounds. Alternator wins iff she can make infinitely many moves.

Let $X$ be a language over the alphabet $A$.  The \emph{$X$-delayed cutting game}, denoted by 
$\mathcal{H}^{X}_\omega(L_1,\dots)$ is similar to a simple cutting game, except that a mini game is played to determine the prefix $p$ and the length $k$ of the match. 
The mini game goes as follows. Firstly, Alternator chooses a tree $t \in X$. Then Constrainer chooses a prefix $p$ of $t$ and a finite ordinal $k$. Finally the two players start to play the simple cutting game $\mathcal{H}^p_k(L_1,\dots,L_k)$. 

When $L_{2i}=L$ and $L_{2i+1}=L^\complement$, then we simply write $\mathcal{H}^p_k(L, L^\complement), \mathcal{H}^p_\infty(L, L^\complement)$ and  $\mathcal{H}^{X}_\omega(L, L^\complement)$.
It was verified in \cite{bp} that a given language $M$ has a Wadge degree less than $\omega$ iff Constrainer has a winning strategy in $\mathcal{H}^\varepsilon_k(M, M^\complement)$, for all but finitely many $k<\omega$.
In \cite{bp} it was also remarked that the language $L$ described in \cite[Section 4.1]{bp} and in Proposition \ref{rem:example} in the Appendix, even  if it is such that Alternator has a winning strategy in every corresponding finite cutting game, she looses the infinite one.
In the next two propositions we establish a link between delayed cutting games and infinite Wadge degrees on the one hand, and infinite simple cutting games and uncountable Wadge degrees on the other hand.
\begin{proposition} (see proof in the Appendix) \label{prop:omega} Let $L$ be a tree language, $[\omega]^+ \leq_W L$ iff Alternator has a winning strategy in $\mathcal{H}^{L}_\omega(L^\complement, L)$. 
\end{proposition}
\begin{proposition} (see proof in the Appendix)  \label{prop:infinity} Let $L$ be a tree language. For every prefix $p$,
$d_W([p]^{-1}L) \geq \omega_1$ iff Alternator has a winning strategy in $\mathcal{H}^p_\infty(L, L^\complement)$. 
\end{proposition}


\section{A characterization of languages of uncountable degree}
\subsubsection*{Games on types and strategy trees.}
Following \cite{bp}, for a given regular language of trees $L$, a prefix $p$ and types $h_i\in H_L$ ($i=1,2,\dots$) we define games on types $\mathcal{H}^p_k(h_1,\dots,h_k)$ and $\mathcal{H}^p_\infty(h_1,h_2,\dots)$. The Constrainer plays as in the simple and infinite cutting games and the task of the Alternator is to play in the $i$--th round a tree of type $h_i$, that is an element of $\alpha_L^{-1}(h_i)$. 

A type tree for $L$ is a tree over the finite alphabet $H_L$. For a given tree $t$, there is a type tree $\sigma_t$ induced by $t$ such that for every node $w \in \dom(\sigma_t)$, 
\begin{equation}
\label{formula:sigmat}
\sigma_t(w)\ \mbox{is the type of the tree}\ t.w.
\end{equation}
Let $\sigma$ be a type tree, and $t$ a tree. We say that a type tree $\sigma$ is \emph{locally consistent} with a tree $t$ if $\dom(\sigma)=\dom(t)$ and for every node $w \in \dom(t)$ such that $t(w)=a$, 
\begin{itemize}
\item if $w$ is a leaf, then $\sigma(w)$ is the type of $a$,
\item if $w$ has two children $m_\ell$ and $m_r$, then $\sigma(w)$ is the type obtained by applying $a$ to the pair $(\sigma(m_\ell), \sigma(m_r))$.
\end{itemize}
\begin{definition} A finite strategy tree is a tuple
$\mathfrak{s}=(t, \sigma_1, \dots, \sigma_k)$ where
\begin{itemize}
\item $t$ is a tree, the support of the strategy and $\sigma_1=\sigma_t$,
\item $\sigma_\ell$ is locally consistent with $t$, for each $\ell \leq k$,
\item for each $w \in \dom(t)$, Alternator has a winning strategy in $\mathcal{H}^\varepsilon_k(\sigma_1(w), \dots, \sigma_k(w))$.
\end{itemize}
An  infinite strategy tree $\mathfrak{s}=(t, \sigma_1, \sigma_2, \dots )$ is defined analogously.
\end{definition}
The root sequence of a strategy tree $\mathfrak{s}=(t, \sigma_1, \sigma_2, \dots )$ is the sequence of types $(\sigma_1(\varepsilon), \sigma_2(\varepsilon), \dots)$.  We define the \emph{alternation} of a sequence $(h_1,\dots,h_\ell)$ of types as the cardinality of the set $\{ i: h_i\neq h_{i+1} \}$. The same definition applies to infinite sequences of types. Let $\mathfrak{s}$ be a finite strategy tree. The \emph{root alternation} of $\mathfrak{s}$ is the alternation of the root sequence, while the \emph{limit alternation} of $\mathfrak{s}$ is the maximal number $k$ such that infinitely many subtrees of $\mathfrak{s}$ have root alternation at least $k$. We say that a set  $\mathfrak{S}$ of finite strategy trees has \emph{bounded root alternation} if there is a  $k$ such that the root alternation of each $\mathfrak{s} \in \mathfrak{S}$ is at most $k$, unbounded otherwise. Analogously for limit alternation.

A finite or infinite strategy tree $\mathfrak{s}=(t, \sigma_1, \dots)$ is \emph{locally optimal} if for every strategy tree $\mathfrak{s}'=(t, \sigma'_1, \dots)$ with same root sequence, and every $i>1$, the depth at which $\sigma_i$ and $\sigma_{i+1}$ first differ is greater than or equal to the depth at which $\sigma'_i$ and $\sigma'_{i+1}$ first differ.
The next Proposition is a very important technical point of \cite{bp}. 
\begin{proposition}[Lemma G.2 in Appendix of \cite{bp}]\label{prop:locality}
 For a regular tree language $L$, if $\mathfrak{S}$ is a set of locally optimal finite strategy trees with both root and unbounded limit alternation, then the strategy graph $G_L$ is recursive.
 \end{proposition}
The next Proposition establishes an important link between infinite cutting games and strategy trees. 
\begin{proposition}\label{prop:types}
 Assume Alternator has a winning strategy in  $\mathcal{H}^\varepsilon_\infty(L, L^\complement)$. Then there is an infinite strategy tree $\mathfrak{s}^\infty$ with infinite root alternation. 
\end{proposition} 
 \begin{proof} 
 Assume Alternator has a winning strategy $f$ in  $\mathcal{H}^\varepsilon_\infty(L, L^\complement)$. The infinite strategy tree $\mathfrak{s}^\infty$ is constructed as follows. First of all, we can represent $f$ as a tree satisfying the following properties:
\begin{itemize}
\item the root is labelled by $\varepsilon$, and its unique child is labelled by Alternator's move obtained by applying the winning strategy $f$ at the first round of the game,
\item if a node $v$ is labelled with a tree $t$, then for every prefix $p$ of $t$ there is a unique child of $v$ labelled by $p$,
\item if a node $v$ is labelled with a prefix $p$, then $v$ has a unique child, and such a child is labelled by the answer obtained by applying the winning strategy $f$ to the position in the cutting game given by the labels of the path from the root to $v$.
\end{itemize}
Notice that nodes at odd depth represent Alternator's moves (according to $f$) and are therefore labelled by trees, while nodes at even depth represent Constrainer's move and are thus labelled by prefixes. From now on, we always identify $f$ and the aforementioned tree.
\begin{claim}\label{claim:strategy}
For every node $v$ of $f$ labelled by a prefix $p$, 
there is an infinite sequence of strategy trees $(\mathfrak{s}^v_\ell: \ell < \omega)$ such that for each $\ell$
\begin{enumerate}
\item $\mathfrak{s}^v_\ell=(t, \sigma_1, \dots, \sigma_\ell)$, with the type $\sigma_{2k+1}(\varepsilon)$ included in $L$ and the type  $\sigma_{2k}(\varepsilon)$ included  $L^\complement$ if $v$ is at depth $2i$ with $i$ even, else dually. In particular this means that  $\sigma_{2k+1}(\varepsilon) \neq \sigma_{2k}(\varepsilon)$;
\item $\mathfrak{s}^n_{\ell+1}$ extends $\mathfrak{s}^v_\ell$, that is  $\mathfrak{s}^v_{\ell+1}=(t, \sigma_1, \dots, \sigma_\ell, \sigma_{\ell+1})$ and $\mathfrak{s}^n_\ell=(t, \sigma_1, \dots, \sigma_\ell)$.
\end{enumerate}
\end{claim}
Given the Claim, from Property 1 we have that
for each node $v$ labelled by a prefix $p$, and each $\ell=1,2,\dots$, $\mathfrak{s}^v_\ell$ has root alternation $\ell$ and defines a winning strategy for Alternator in $\mathcal{H}^p_\ell(L, L^\complement)$ if $v$ is at depth $2i$ with $i$ even, in $\mathcal{H}^p_\ell(L^\complement, L)$ otherwise. 
Let \[ \mathfrak{s}^\varepsilon_\ell = (t, \sigma_1, \dots, \sigma_\ell)\ \mbox{for}\ \ell=1,2,\dots.\] The required   
infinite stategy tree is defined as 
$ \mathfrak{s}^\infty = (t, \sigma_1, \dots )$.
It remains to prove the Claim.
Firstly, by induction with respect to $\ell=1,2,\dots$ we will assign a strategy tree $\mathfrak{s}^v_\ell$ to each node $v$ of $f$ labelled by a prefix. 
In the process of inductive construction we will also verify that Property 1 of the Claim is satisfied. Verification of Property 2 will be done later. Let us start from a remark that given an infinite sequence of type trees $(\sigma_1, \dots)$, by compactness there is a converging subsequence $(\sigma'_1, \dots)$. We assume that every time we have to choose  
a converging subsequence $(\sigma'_1, \dots)$ of a given sequence $(\sigma_1, \dots)$, we always choose the same subsequence and denote it's limit as 
$\limit(\sigma_1, \dots)$. We also assume that given a tree $t$, we have fixed an enumeration $(p_1, \dots)$ of all its prefixes such 
that sequence $(p_k)_{k=1,2,\dots}$ converge to the tree $t$.
For $\ell=1$, 
it is enough to take for each node $v$  \[ \mathfrak{s}^v_1=(t, \sigma_t),\] where $t$ is given by applying $f$ to the considered position and $\sigma_t$ is defined by formula (\ref{formula:sigmat}) at the beginning of this section. By choice of $\sigma_t$, Property 1 is satisfied. For $\ell>1$ we proceed as follows. We assume the construction performed for $\ell-1$. Fix any node $v$ labelled by a prefix $p$. Assume that a tree $t$ is the answer given by $f$  at the position in the game given by the path from the root to the node $v$. To every prefix $p$ of $t$ corresponds a child $w$ of $v$ to which we already associated a strategy tree $\mathfrak{s}^w_{\ell-1}=(t^p, \sigma^p_{2}, \dots, \sigma^p_{\ell})$. 
Let us thence consider the sequence $(p_1, \dots)$, with limit $t$ and the sequences
$(t^{p_1}, \dots)$, $(\sigma^{p_1}_2\dots)$, $\dots,$  $(\sigma^{p_1}_{\ell}\dots)$.  The limits $\limit(\sigma^{p_k}_2),\dots,
\limit(\sigma^{p_k}_{\ell})$ were chosen in advance and are equal $\sigma^*_2, \dots, \sigma^*_\ell$. Since each $t^{p_k}$ extends $p_k$, the limit $t^*$ of  $(t^{p_1},t^{p_2},\dots)$ is $t$.
Now, for each $p$, the type trees $(\sigma^p_{2}, \dots, \sigma^p_{\ell})$
are locally consistent with $t^p$. Furthermore, given a sequence of trees $(t_1, \dots)$  that converges to $t^*$ and a sequence of type trees $(\sigma_1, \dots)$  that converges to $\sigma^*$, if $\sigma_k$ is locally consistent with $t_k$ for every k, then $\sigma^*$ is locally consistent with $t^*$.
From this fact follows that the limits $\sigma^*_2, \dots, \sigma^*_k$ are locally consistent with $t$. Finally, define $\sigma^*_1$ to be $\sigma_t$ as in formula (\ref{formula:sigmat}). We have just proved that
$\mathfrak{s}^v_\ell = (t, \sigma^*_1, \dots, \sigma^*_\ell)$
is a strategy tree. From induction hypothesis together with definition of $\sigma_t$ and preservation of Property 1 under limits follows that $\mathfrak{s}^v_\ell$ also 
satisfies Property 1. 

We now verify that the described procedure preserves Property 2. For $\ell=1$ there is nothing to check. For the induction step, we reason as follows. Assume the Property holds for each node and for each $k<\ell$. Now, let us consider an arbitrary node $v$. We have to prove that $\mathfrak{s}^v_{\ell+1}$ extends $\mathfrak{s}^v_{\ell}$. 
By induction hypothesis,  $\mathfrak{s}^w_{\ell-1}=(t^p, \sigma^p_{2}, \dots, \sigma^p_{\ell})$ and
$\mathfrak{s}^w_{\ell}=(t^p, \sigma^p_{2}, \dots, \sigma^p_{\ell}, \sigma^p_{\ell+1})$, for every node $w$  in the described procedure. Since the limits have been fixed in advance, we have that $\mathfrak{s}^w_{\ell}=(t, \sigma_t, \sigma^*_2, \dots, \sigma^*_\ell)$ and $\mathfrak{s}^v_{\ell+1}=(t, \sigma_t, \sigma^*_2, \dots, \sigma^*_\ell, \sigma^*_{\ell+1})$, meaning that the latter extends the former. This concludes the proof of the Claim.
 \end{proof}
 
Using the above Proposition, we can generalize to infinite games Proposition 5.2 from \cite{bp}:
\begin{proposition}\label{prop:tree_to_types}
For a regular language $L$ 
the following conditions are equivalent.
 \begin{enumerate}
\item Alternator wins the game $\mathcal{H}^\varepsilon_\infty(L, L^\complement)$, 
 \item  There are tree types $h, g\in H_L$, such that $h\neq g$ and Alternator wins $\mathcal{H}^\varepsilon_\infty(h, g)$.
 \end{enumerate}
\end{proposition}
The proof of Proposition \ref{prop:tree_to_types} can be found in the Appendix.
We will use the following Lemma, presented in \cite{bp} for finite strategy trees, with proof extending straightforwardly to infinite strategy trees.
\begin{lemma}\label{lemma:locallyoptimal}
For every finite or infinite strategy tree, there is a locally optimal strategy tree with same root sequence.
\end{lemma}
The next Lemma  follows immediately from the definition of a strategy tree.
\begin{lemma}\label{lemma:short_strategy}
Let $\mathfrak{s}=(t, \sigma_1, \dots, \sigma_\ell)$ be a strategy tree. For the game $\mathcal{H}^\varepsilon_\ell(\sigma_1(\varepsilon), \dots, \sigma_\ell(\varepsilon))$ and a strategy of Constrainer given by always cutting at level $i$, Alternator wins  by playing as follows:
\begin{itemize}
\item at first, Alternator plays $t$, then
\item for each port $w$ at level $i$ of the multi context given by Constrainer's move, Alternator plugs in the tree given by her winning strategy $\mathcal{H}^\varepsilon_\ell(\sigma_1(w), \dots, \sigma_\ell(w))$.
\end{itemize}
In particular, if from a certain $j<\ell$ on $\sigma_k(w)=\sigma_{k+1}(w)$, $j\leq k < \ell$, then for each round $k$ such that $j< k < \ell$ Alternator always plugs in the same tree of type $\sigma_j(w)$ chosen at round $j$.
\end{lemma}
\subsubsection*{An Effective Characterization.}
\label{sec:the main result}
Everything now is ready to prove the main result of this paper.
\begin{theorem}\label{theorem:main}
Let $L$ be a regular tree language given by a non-deterministic tree automaton $\A$. The following conditions are equivalent:
\begin{enumerate}
\item The strategy graph $G_L$ is recursive.
\item $d_W(L) \geq \omega_1$
\end{enumerate}
In particular, since the graph $G_L$ is computable from the automaton $\A$, it is decidable whether the language accepted by $\A$ is of Wadge degree greater than or equal to $\omega_1$.  
\end{theorem}
\begin{proof}
${\bf (1) \Rightarrow (2).}$  Assume the strategy graph is recursive. This means that there exists a strongly connected component that contains two nodes $(v, h)$ and $(v', h')$ with $h \neq h'$. 
Thanks to Proposition \ref{proposition:path to edge}, if there exists a path between $(v, h)$ and $(v', h')$, there is also an edge between $(v, h)$ and $(v', h')$. 
Moreover, for vertices $(v_1,h_1),(v_2,h_2),\dots$, if for every $i=1,2,\dots$ there is an edge from $(v_i, h_i)$ to $(v_{i+1}, h_{i+1})$, this means that Alternator has a winning strategy in $\mathcal{H}^\varepsilon_\infty(h_1,h_2,\dots)$. So, take $(v_i, h_i)= (v,h)$ for $i$ even, and $(v_i, h_i)= (v',h')$ for $i$ odd. This shows that Alternator has a winning strategy in $\mathcal{H}^\varepsilon_\infty(h, h')$. By Proposition \ref{prop:tree_to_types}
Alternator has a winning strategy in  $\mathcal{H}^\varepsilon_\infty(L, L^\complement)$.

${\bf (2) \Rightarrow (1).}$ 
By Propositions \ref{prop:infinity}  and \ref{prop:locality}, it is enough to verify that 
 if Alternator has a winning strategy in $\mathcal{H}^\varepsilon_\infty(L, L^\complement)$ then there is a set $\mathfrak{S}$ of locally optimal finite strategy trees  with both root and limit unbounded alternation. 
Assume Alternator has a winning strategy $f$ in  $\mathcal{H}^\varepsilon_\infty(L, L^\complement)$. From Proposition \ref{prop:types} there is a strategy tree $\mathfrak{s}^\infty = (t,\sigma_1,\dots) $ with infinite root alternation. 
 By Lemma \ref{lemma:locallyoptimal} we can assume that $\mathfrak{s}^\infty$ is  locally optimal. Let us define 
\[ \mathfrak{S}= \{ (t,\sigma_1,\dots,\sigma_k): k=1,2,\dots\}. \]
Note that each element of $\mathfrak{S}$ is locally optimal.
Now, assume limit alternation of $\mathfrak{S}$ is bounded.
From this fact and since every element of $\mathfrak{S}$ is a prefix of $\mathfrak{s}^\infty$, it holds that with respect to $\mathfrak{s}^\infty$, the set of subtrees of $t$ with infinite root alternation has to be finite. This means that $\mathfrak{s}^\infty$ satisfies the following property:
\begin{description}
\item[(*)] 
there is a finite set $X$ of nodes of $t$ satisfying the following properties:
\begin{itemize}
\item the root is included in $X$, and each node of $X$ is at most at depth $i$ in $t$, 
\item $\sigma_k(v)\neq \sigma_{k'}(v)$, for every node $v$ in the set $X$, and $\sigma_k(w)= \sigma_{k'}(w)$, for every node $w$ of $t$ of depth $i+1$, for some $k, k'$ , with $k < k' \leq j$.
\end{itemize}
\end{description}
The strategy tree $\mathfrak{s}=(t, \sigma_1, \dots, \sigma_j)$ from $\mathfrak{S}$, where $j$ is given by the previous property, also satisfies the property (*) above (for the same $X$ and the same $k, k'$).


Let us consider the game $\mathcal{H}^\varepsilon_j( \sigma_1(\varepsilon), \dots, \sigma_j(\varepsilon))$ where at first Alternator plays $t$ and then Constrainer uses the strategy given by cutting always at level $i+1$. We can therefore apply Lemma \ref{lemma:short_strategy} and assume that Alternator plays the winning strategy described there. This implies that the trees played at round $k$ and $k'$ are the same, say $t'$ (from the root to level $i$ they are the same, because the Constrainer insists on this and below they are the same, because the Alternator plays the same answers in rounds $k$ and $k'$). But by local consistency, since $\sigma_k(\varepsilon)\neq \sigma_{k'}(\varepsilon)$, the two trees should have two different types, a contradiction. We therefore conclude that limit alternation of $\mathfrak{S}$ is unbounded. The method of proof is illustrated by Figure \ref{figure:consistency} in the Appendix .
\end{proof}


\section{Conclusion}
The algorithm provided in \cite{bp} decides whether a given non-deterministic automaton  accepts a language which is a Boolean combination of open sets or equivalently is of a  Wadge degree smaller than $\omega$. 
By the same approach we showed an algorithm which decides whether a given non-deterministic automaton accepts a language in ${\bf \Delta^0_2}$ or equivalently, a language of a Wadge degree smaller than $\omega_1$. 
We propose for further research the following three generalizations of the result presented in this paper:

\noindent
{\bf 1}. 
For a given $n=1,2,\dots$ there are natural topological games which characterize languages of Wadge degrees smaller than $\omega^n$. Moreover, there are known examples of regular languages of  degree $\omega^n$. It would be a desirable and perhaps more involved extension of results in \cite{bp} if for a given $n$ one can provide an algorithm deciding whether a given non-deterministic automaton accepts a language of degree smaller than $\omega^n$. 

\noindent
{\bf 2.} In the absence of examples of regular languages between Wadge degree $\omega^\omega$ and Wadge degree $\omega_1$, one could reasonably expect, that the decidability result in the present paper should show that indeed any regular language of countable Wadge degree is of Wadge degree smaller than $\omega^\omega$. However, this question still remains open.  

\noindent
{\bf 3}. Regarding higher Borel classes, in particular regular languages which are Boolean combinations of ${\bf \Sigma^0_2}$ sets, the following extension of the method in \cite{bp} seems to be plausible. The cutting game is based around restrictions of moves by prefixes, that is 
by languages in ${\bf \Delta^0_1}$. Its  topological counterpart on the next Borel level is a game, where the Constrainer is allowed to play constraints which are regular languages in ${\bf \Delta^0_2}$. This leads to a natural topological characterization similar to the results in Section \ref{section:games}, but the algebraic counterpart of this generalized cutting game is not yet fully understood.



\clearpage
\section*{Appendix}

\subsection*{Set-Theoretical Operations and Wadge degrees}

In this part of the Appendix, we follow the expositions in \cite{dup1,dm7} and show how to generate canonical sets complete for each $\equiv_W$-class whose corresponding Wadge degree is countable. We assume that the alphabet $A$ has at least two elements.

\vspace{0.2cm}
\noindent {\bf Sum :}
\hspace{0.1cm}

\noindent  Given two languages $L$ and $M$ over $A$, we define
 the set $L \to M$ as the set of trees $t$ over $A \cup\{a\}$, with $a \notin A$, satisfying one of the following conditions:

\begin{itemize}
\item $t.0 \in L$ and $a = t(10^n)$ for all $n$,
\item $10^{n}$ is the first node on the path $10^*$ such that $a \neq t(10^{n})$ and $t.10^{n}0 \in M$.
\end{itemize}

Based on this operation, we thus define the sum operation.
Let $L$ and $ M $ be two languages over $A$. The set $M+ L$ is defined as $L \to M^\pm$. 

Let us a provide an intuition behind this construction: in a Wadge game $\mathcal{W}(M+L, X)$ Player I plays like in a Wadge game $\mathcal{W}(L, X)$, but in addition at any moment of the play Player I may decide to erase everything played so far and start a game
$\mathcal{W}(M^\pm, X)$. 

%

\vspace{0.2cm}
\noindent {\bf Countable multiplication :} \hspace{0.1cm}

\noindent  Let $\kappa$ be a countable ordinal, and let  $L_\alpha$ be a language over $A$,
for every $\alpha < \kappa$. Fix any bijection $f: \omega \to \kappa$. Thus, the language $\mathrm{sup}^-_{\alpha<\kappa} L_\alpha$ is defined as the set of trees $t$ over $A\cup\{b\}$ satisfying the following conditions for some $k$:
\begin{itemize}
\item $0^k$ is the first node on $0^*$ labeled with $b$,
\item $t0^k1 \in L_{f(k)}$.
\end{itemize}

Define also
$\sup^+ _{\alpha < \kappa} L_\alpha$ as $\sup^-_{\alpha < \kappa} L_\alpha\cup \{t :\, \forall_n\; t(1^n)\neq b\}$.
The difference from the previous operation is that now, when the
Player does not plays $b$ on the leftmost branch, the obtained tree is
accepted. Note that the operations are dual. 
%

The set-theoretic counterpart of the countable multiplication is thus  inductively defined as follows. 
Let $L$ be a language: 
\begin{itemize}
\item $L \bullet 1 = L$, 
\item $L \bullet (\alpha + 1) = (L \bullet \alpha)+L$, 
\item $L \bullet \kappa = \sup^+_{\alpha < \kappa} L \bullet \alpha$ when $\kappa$  is some limit countable ordinal.
\end{itemize}

Let us a provide an intuition behind this construction:  in a Wadge game $\mathcal{W}(L\bullet \kappa, X)$ Player I plays like in a Wadge game $\mathcal{W}(L, X)$, but in addition at any moment of the play Player I may decide to erase everything played so far and start either a game $\mathcal{W}(L, X)$ or a game $\mathcal{W}(L^\complement, X)$. 
With every such change Player I decreases the ordinal $\kappa$. Hence the aforementioned procedure is producing a decreasing finite sequence of ordinals below $\kappa$ and preventing Player I from reinitializing the game indefinitely.

Finally, we remark that  the defined set-theoretical operations are the counterpart of the corresponding ordinal operations on Wadge degrees.

\begin{lemma}[\cite{dup1}]
\label{r_mult}
Let $L$ and $ M$ be two non self dual languages. Then 
\begin{itemize}
\item $d_W(L+M)=d_W(L)+d_W(M)$, 
\item$d_W(\sup^+ _{\alpha < \kappa} L_\alpha)= d_W(\sup^- _{\alpha < \kappa} L_\alpha)=\sup_{\alpha < \kappa}d_W( L_\alpha)$.
\item $d_W(L \bullet \kappa)= d_W(L) \cdot \kappa$, for every countable ordinal $\kappa$.
\end{itemize}
\end{lemma}

The sign of the degree of a non self dual set Wadge equivalent 
to $\sup^+ _{\alpha < \kappa} L_\alpha$, for some   family $(L_\alpha: \alpha < \kappa)$, is $+$, dually for the operation $\sup^-$. 
For each $\alpha < \omega_1$, and sign $\epsilon \in \{+,-,\pm\}$, we use  $[\alpha]^\epsilon$ to also denote the canonical complete language of signed Wadge degree $[\alpha]^\epsilon$ whose construction is given by the previous operations.

For each $n>1$, every ${\bf \Sigma}^0_n$-complete set has signed Wadge degree $[\exp^{n-1}(1)]^-$, and every ${\bf \Pi}^0_n$-complete set has thus signed degree $[\exp^{n-1}(1)]^+$, where $\exp(\alpha) = \omega_1^\alpha$ and $\exp^{k+1}(\alpha) = \omega_1^{\exp^k{\alpha}}$.

\begin{proposition}[Section 4.1 in \cite{bp}]
\label{rem:example}
Let $L\subset T_{\{a,b\}}$ be the set of trees $t$ such that for some $n$, $0^n$ is a leaf, $t(0^k)=a$ and for each $k\in \{1, \dots, n\}$ the tree $t.0^k1$ is either finite or contains no $b$. Then $L$ has Wadge degree $[\omega]^-$. 
\end{proposition}

\begin{proof} We have to prove that $[\omega]^-\leq_W L$ and $L\leq [\omega]^-$. Consider first the Wadge game $\mathcal{W}([\omega]^-, L)$. We first show that 
Player II has a winning strategy in this game and then Lemma \ref{lemma:wadge} will imply $[\omega]^-\leq_W L$. The informal argument goes as follows. As long as Player I plays rejecting, she plays a complete binary tree labelled with $a$. Now, assume that Player I stops player rejecting, she decreases her ordinal from $\omega$ to $n$ and starts playing rejecting. Then Player II stops the branch plays the node $0^{n+1}$ as being a leaf of her final tree, she plays $t(010^{n-1})=b$, and she stops playing in any branch of the subtree of $t.1$, except for $01^\omega$. Assume that Player I moves to $n-1$ and starts playing accepting, then Player 2 stops the branch on $01^\omega$. Since she thus played a finite tree in $t.01$, she is now playing accepting. By following this strategy on the successive subtree $t.01^i$, Player II can follow the moves of Player I and thus win the game.

For the other direction, we have to show that Player II has a winning strategy in the game $\mathcal{W}(L,[\omega]^-)$. We notice the following, which actually was implicit in the previous description of the winning strategy for Player II. The best Player I in $\mathcal{W}(L,[\omega]^-)$ can do is to apply the following strategy. As long as she plays $t(0^k) = a$, she is rejecting. As soon as she kills this branch, we have to look at what she was playing in each subtrees  $t.0^k1$. Consider an arbitrary such subtree. As long as she plays only nodes labelled by $a$ she is accepting in this subtree. But as soon as she plays a $b$ and keeps active a branch in this subtree, she is globally rejecting. To be anew accepting she has to kill all active branches. Once she has killed all active branches of a subtree $t.0^k1$, she cannot change the status of the considered subtree. This means that as soon as she has fixed the length $k$ of the branch $0^\omega$ she wants to play, she can  alternate at most $2k+1$ times between being accepting and being rejecting in her play in the game. The winning strategy for Player II in $\mathcal{W}(L, [\omega]^-)$ is thus just to play rejecting, to wait the length $k$ of the branch, and then to decrease $\omega$ to $2k+1$ and follow the behavior of Player I.
\end{proof}

\vspace{0.2cm}
\noindent {\bf The level $\omega_1$ :}
\hspace{0.1cm}

\noindent We present here two canonical non self dual languages of Wadge degree $\omega_1$. The first is the set of tree $t$ such that there is a $n<\omega$ such that $t.0^n1 \in [3]^-$. This set is ${\bf \Sigma}^0_2$-complete and is thus of signed Wadge degree $[\omega_1]^-$. Its ${\bf \Pi}^0_2$-complete complement of signed degree $[\omega_1]^+$ is the language of all trees $t$ such that for every $n<\omega$, the subtree $t.0^n1$ is in  $[3]^+$.

From the perspective of a Player in a Wadge game, a Player in charge of $[\omega_1]^-$ is like a Player starting to play rejecting, that is being in charge of $\emptyset$, with at every point the possibility of reinitializing anew the play and being in charge of its complement, and so on, with the condition that if she restarts the game infinitely often, at the end of the game she is rejecting. 
The dual description holds for a Player in charge of $[\omega_1]^+$.

\subsection*{Proof of Proposition \ref{prop:omega}}
For the direction from left to right, we reason as follows. Let $f$ be the winning strategy for Player II in $\mathcal{G}([\omega]^+, L)$. As a first move, Alternator plays the tree $t$ given by applying the strategy $f$ against Player I in a Wadge game where she is playing always accepting. Now, suppose that Constrainer plays a prefix $p$ of depth $\ell$ and a number $k$ (this is the mini-game from the definition of delayed cutting game). Thus Alternator looks at the shadow Wadge game used to determine $t$, but at $k+1$ round, she makes Player I erasing the game and play a new Wadge game into the set $[k']^-$, for a $k' \geq k + \ell +1$. She then  applies her winning strategy $f$ in such a game where she makes Player I playing rejecting. Assume the obtained tree is $t'$. We have that:
\begin{itemize}
\item $t' \notin L$ and $t'$ extends $p$,
\end{itemize}
thus the one described is an admissible move for Alternator.
Now, assume at next round Constrainer chooses a prefix $p_1$ (without loss of generality extending $p$) of $t'$, whose depth is $\ell_1$. Then in the shadow Wadge match, Alternator modifies Player I 's strategy as follows: at turn $k_1 \geq k' + \ell_1$ she decreases the ordinal and starts to play accepting. The obtained tree by applying the winning strategy $f$ to the play when  Player  I keeps playing accepting is next Alternator 's moves. For the same reasons as before, such a move is admissible. By continuing such a strategy, it is clear that Alternator wins.

\begin{figure}
\begin{center}
\includegraphics[height=3in]{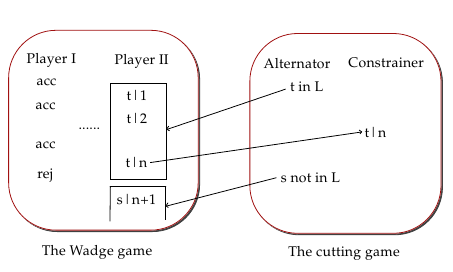}
\caption{The flow of information between games $\mathcal{H}^{L}_\omega(L^\complement, L)$ and $\mathcal{W}([\omega]^+, L)$. The figure illustrates proof of right to left implication in Proposition \ref{prop:omega}. 
\label{myfig1} }
\end{center}
\end{figure}

For the direction from right to left, we describe a winning strategy for Player II in $\mathcal{G}([\omega]^+, L)$, see Figure \ref{myfig1}. In the back she keeps track of a shadow match in the game $\mathcal{H}^{L}_\omega(L^\complement, L)$ where she applies the  winning strategy for Alternator. 
As long as Player I plays accepting, Player II just plays the initial choice of Alternator. Now, assume that at round $n$, Player II decides to decrease his ordinal to $k$ and to play rejecting. Then, in the shadow match, Player II makes Constrainer play a prefix of depth $n-1$ and the ordinal $k+1$. She then looks at Alternator winning move, which is a certain tree $t_1$. Player II starts thence to play into $t_1$. By construction $t_1$ is not in $L$, meaning that if Player I continues to play rejecting, she wins. Assume that at round $n_1> n$ Player I decreases his ordinal of one and decides to start to play accepting, then in the second round of the shadow match, Player II forces Constrainer to choose a prefix of depth $n_1$, and Alternator to apply her winning strategy to obtain a tree $t_2$ extending the chosen prefix of $t_1$. Player II therefore just start to play into $t_2$. Since this tree is in $L$, as before if Player I keeps playing accepting, she looses. Player II wins the game if uses repeatedly this strategy. 

\subsection*{Proof of Proposition \ref{prop:infinity}}

For the direction from left to right, it is enough to apply the fact that 
Player II has a winning strategy in $\mathcal{G}(M, [p]^{-1}L)$, where $M$ is either $[\omega_1]^+$ or $[\omega_1]^-$. 
%
The other direction is verified by showing that for every countable ordinal $\kappa$, Player II has a winning strategy in $\mathcal{G}([\kappa]^+, [p]^{-1}L)$ and $\mathcal{G}([\kappa]^-, [p]^{-1}L)$.
This is done by induction. 
If $\kappa=1$, then the claim is trivially proved.
Assume now that $\kappa= \beta + 1$. We only consider the case for $+$, the case for $-$ being immediate  by considering the winning strategy for Alternator in $\mathcal{H}^{p'}_\infty(L^\complement, L)$, for some prefix $p'$ of the first winning move for Alternator in $\mathcal{H}^p_\infty(L, L^\complement)$. At first Player II applies the induced winning strategy in $\mathcal{G}([1]^+, [p]^{-1}L)$. Now assume that at round $n$ Player I decides the erase everything and being in charge of $[\beta]^-$ (the case for $[\beta]^+$ is exactly the same). Assume that before her turn at round $n$, Player II has played $p'$. From round $n+1$ she just apply the winning strategy given by the induction hypothesis in $\mathcal{G}([\beta]^-, [p']^{-1}L)$. 
We now verify the limit case $\kappa= \beta\cdot\omega$. As before, we only consider the case for $+$. Player II applies the induced winning strategy in $\mathcal{G}([1]^+, [p]^{-1}L)$. Assume that at round $n$ Player I decides to move everything and being in charge of $[\lambda]^\varepsilon$, for some $\lambda < \kappa$ and $\varepsilon\in \{+,-\}$. Then from round $n+1$ Player II just apply the winning strategy given by the induction hypothesis in $\mathcal{G}([\lambda]^\varepsilon, [p']^{-1}L)$, where $p'$ is her position after round $n$. 

\subsection*{Proof of Proposition \ref{prop:tree_to_types}}

The direction from (1) to (2) is an immediate corollary of Proposition \ref{prop:types}.
For the direction from (2) to (1) we reason as follows. Since $g$ and $h$ are different elements of the syntactic algebra, it follows that there must be some multi-context $c$ such that the tree type $c[g]$ is contained in $L$, while the tree type $c[h]$ is disjoint with $L$. 
Now, if Alternator has a winning strategy in $\mathcal{H}^\varepsilon_\infty(g, h)$, then she has a winning strategy in $\mathcal{H}^\varepsilon_\infty(c[g], c[h])$, for every multi context $c$. 
This implies that Alternator has a winning strategy in the game $\mathcal{H}^\varepsilon_\infty(L, L^\complement)$.

\subsection*{Proof of Lemma \ref{lemma:locallyoptimal}}
Let $\mathfrak{s}=(t, \sigma_1, \sigma_2, \dots)$ be a strategy tree. We construct the locally optimal strategy tree $\mathfrak{s}'=(t, \sigma_1', \sigma_2', \dots)$ by induction as follows. We put first $\sigma'_1=\sigma_1$. Then, consider the set of all strategy trees (finite or infinite) that are locally consistent with $t$ and which have the same root value as $\sigma_2$. This set is a closed set, and therefore is compact. It follows that some element of this set minimizes the distance with respect to $\sigma'_1$. Such element will be the new $\sigma'_2$. We proceed likewise for next coordinates. 

\subsection*{A figure ilustrating the proof of Theorem \ref{theorem:main}}
\begin{figure}
\begin{center}
\includegraphics[height=2.5in]{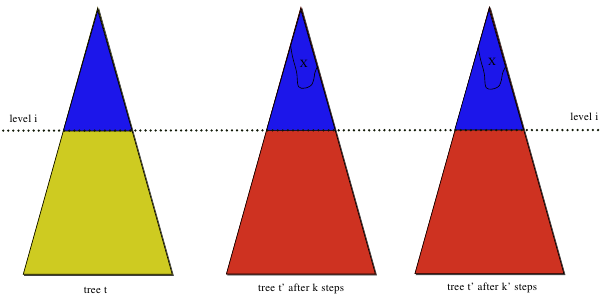}
\caption{Local consistency and bounded limit alternation would force bounded root alternation in Theorem \ref{theorem:main}. 
\label{figure:consistency}} 
\end{center}
\end{figure}


\begin{thebibliography}{00}

{\footnotesize
\bibitem{blumensath}
\textsc{Blumensath} A., \emph{An algebraic proof of Rabin's theorem}, Theoretical Computer Science \textsc{478}, pp. 1--21, 2013.

\bibitem{bp}
\textsc{Boja\'nczyk} M., \textsc{Place} T., \emph{Regular Languages of Infinite Trees That Are Boolean Combinations of Open Sets}, ICALP (2) \textsc{2012}, 104--115. Version with Appendix available at {\tt http://www.mimuw.edu.pl/$\sim$bojan/papers/bool-open.pdf}

\bibitem{bojtrees}
\textsc{Boja\'nczyk} M., \emph{Algebra for trees}, In Handbook of Automata Theory, European
Mathematical Society Publishing House, to appear.

\bibitem{bojidziaszek}
\textsc{Boja\'nczyk} M., \textsc{Idziaszek} I., \emph{Algebra for infinite forests with an
application to the temporal logic EF}, CONCUR \textsc{2009}, 131--145.


\bibitem{cklvb}
\textsc{Colcombet} T., \textsc{Kuperberg} D., \textsc{Löding} C., and \textsc{Vanden Boom} M., \emph{Deciding the weak definability of Büchi definable tree languages}, CSL \textsc{2013}, 215--230.


\bibitem{dup1}
\textsc{Duparc} J., \emph{Wadge Hierarchy and Veblen Hierarchy Part 1: Borel Sets of Finite Rank}, Journal of Symbolic Logic \textsc{66} (1), 56--86, 2001.



\bibitem{dm7}
\textsc{Duparc}, J., \textsc{Murlak} F., \emph{On the Topological Complexity of Weakly Recognizable Tree Languages}, FCT \textsc{2007}, 261--273.

\bibitem{fms}
\textsc{Facchini} A., \textsc{Murlak} F., \textsc{Skrzypczak} M.,  \emph{Rabin-Mostowski index problem: a step beyond deterministic automata}, LICS \textsc{2013}, 499--508.


\bibitem{kechris}
\textsc{Kechris} A., \emph{Classical Descriptive Set Theory}, Springer, \textsc{1995}.


\bibitem{martin} 
\textsc{Martin} D.A., \emph{Borel determinacy}, The Annals of Mathematics, \textsc{102} (1975), 363--371.


\bibitem{murlak2}
\textsc{Murlak} F., \emph{On deciding topological classes of deterministic tree languages}, CSL \textsc{2005}, 428--442. 

\bibitem{murlak}
\textsc{Murlak} F., \emph{The Wadge Hierarchy of Deterministic Tree Languages}, ICALP \textsc{2006}, 408--419. 

\bibitem{niwalu}
\textsc{Niwiński} D. and \textsc{Walukiewicz} I., \emph{A gap property of deterministic tree languages}, Theoretical Comput. Sci. \textsc{303} (2003), 215--231.



\bibitem{pinperrin}
\textsc{Perrin} D, \textsc{Pin} J.E., \emph{Infinite Words: Automata, Semigroups, Logic and Games}, Academic Press, \textsc{2004}.

\bibitem{rabin}
\textsc{Rabin} M.O., \emph{Weakly definable relations and special automata}, in Foundations of Set Theory, Y. Bar-Hillel ed., \textsc{1970}, 1--23.

\bibitem{wadge}
\textsc{Wadge} W.W., \emph{Reducibility and Determinateness on the Baire Space}, Ph.D. Thesis, Berkeley, \textsc{1984}.

\bibitem{wagner}
\textsc{Wagner} K., \emph{On $\omega$-regular sets}, Inform.~and Control, \textsc{43}, 123--177, 1979. }

\end{thebibliography}
\end{document}